\documentclass[%
 reprint,
superscriptaddress,
 amsmath,amssymb,
 aps,
 prab,
]{revtex4-2}

\usepackage{graphicx}
\usepackage{dcolumn}
\usepackage{bm}
\usepackage{xcolor} 

\begin{document}

\preprint{APS/123-QED}

\title{Demonstration of ultra-low emittance beams in a kHz laser wakefield accelerator and their application to electron diffraction}

\author{J. Monzac}
\email{josephine.monzac@laplace.univ-tlse.fr}
\affiliation{Laboratoire d’Optique Appliquée (LOA), CNRS, École polytechnique, ENSTA, Institut Polytechnique de Paris, Palaiseau, France}
\author{S. Smartsev}
\affiliation{Laboratoire d’Optique Appliquée (LOA), CNRS, École polytechnique, ENSTA, Institut Polytechnique de Paris, Palaiseau, France}
\author{J. Huijts}
\affiliation{Laboratoire d’Optique Appliquée (LOA), CNRS, École polytechnique, ENSTA, Institut Polytechnique de Paris, Palaiseau, France}
 \author{A. Vernier}
\affiliation{Laboratoire d’Optique Appliquée (LOA), CNRS, École polytechnique, ENSTA, Institut Polytechnique de Paris, Palaiseau, France}
 \author{I. A. Andriyash}
\affiliation{Laboratoire d’Optique Appliquée (LOA), CNRS, École polytechnique, ENSTA, Institut Polytechnique de Paris, Palaiseau, France}
\author{V. Tomkus}
\affiliation{Center for Physical Sciences and Technology, Savanoriu Ave. 231, LT-02300, Vilnius, Lithuania}
\author{V. Girdauskas}
\affiliation{Center for Physical Sciences and Technology, Savanoriu Ave. 231, LT-02300, Vilnius, Lithuania}
\affiliation{Vytautas Magnus University, K. Donelaicio St. 58. LT-44248, Kaunas, Lithuania}
 \author{G. Raciukaitis}
 \affiliation{Center for Physical Sciences and Technology, Savanoriu Ave. 231, LT-02300, Vilnius, Lithuania}
 \author{M. Mackevi\v ci\=ut\.{e}}
\affiliation{Center for Physical Sciences and Technology, Savanoriu Ave. 231, LT-02300, Vilnius, Lithuania}
\author{V. Stankevic}
\affiliation{Center for Physical Sciences and Technology, Savanoriu Ave. 231, LT-02300, Vilnius, Lithuania}
  \author{A. Cavagna}
\affiliation{Laboratoire d’Optique Appliquée (LOA), CNRS, École polytechnique, ENSTA, Institut Polytechnique de Paris, Palaiseau, France}
 \author{J. Kaur}
\affiliation{Laboratoire d’Optique Appliquée (LOA), CNRS, École polytechnique, ENSTA, Institut Polytechnique de Paris, Palaiseau, France}
 \author{A. Kalouguine}
\affiliation{Laboratoire d’Optique Appliquée (LOA), CNRS, École polytechnique, ENSTA, Institut Polytechnique de Paris, Palaiseau, France}
\author{R. Lopez-Martens}
\affiliation{Laboratoire d’Optique Appliquée (LOA), CNRS, École polytechnique, ENSTA, Institut Polytechnique de Paris, Palaiseau, France}
\author{J. Faure}
\affiliation{Laboratoire d’Optique Appliquée (LOA), CNRS, École polytechnique, ENSTA, Institut Polytechnique de Paris, Palaiseau, France}

\date{\today}

\begin{abstract}
We present a compact, cost-effective method for measuring the emittance of kHz-repetition-rate laser-wakefield accelerated electron beams using a permanent solenoid. The measured normalized emittance, $\epsilon_n = 124\,\mathrm{nm \cdot rad}$ ($\simeq 0.04 \pi\,\mathrm{mm \cdot mrad}$) at $2.7\,$MeV, is comparable to that of ultra-low emittance radiofrequency guns used for electron diffraction. Leveraging this low emittance, we successfully applied the electron beam to electron diffraction. We demonstrate diffraction images obtained from a single-crystal silicon nanomembrane sample, clearly resolving diffraction peaks across multiple orders. 
\end{abstract}

\maketitle

\section{\label{sec1} Introduction}

Laser wakefield acceleration (LWFA) is a promising method for accelerating electrons to relativistic energies over very short distances using high-intensity laser pulses to drive plasma wakefields \cite{tajima_laser_1979}. Modern LWFA systems, driven by $100\,$TW to PW lasers, routinely produce electron beams with energies ranging from a few MeV to several GeV within centimeters \cite{monzac_optical_2024, salehi_laser-accelerated_2021, kim_stable_2017, oubrerie_controlled_2022, miao_multi-gev_2022, gonsalves_petawatt_2019, aniculaesei_acceleration_2023}. MeV-scale electron beams accelerated via LWFA hold significant potential for Ultrafast Electron Diffraction (UED), a pump-probe technique that probes ultrafast structural dynamics in materials \cite{miller_femtosecond_2014, zewail_4D_2009, sciaini_femtosecond_2011}. In UED, a laser pulse excites the material, and electron diffraction images are captured at varying time delays to observe its time-resolved response. The UED's temporal resolution is thus limited by the probe beam duration, space charge, and shot-to-shot jitter \cite{sciaini_femtosecond_2011, van_Oudheusden_compression_2010, musumeci_laser-induced_2010}. LWFA beams, however, offer intrinsic advantages: femtosecond duration \cite{lundh_few_2011}, high charge, and minimal timing jitter (the accelerating structure is directly driven by the laser), and could enable sub-10-fs resolution \cite{faure_concept_2016}. For UED applications, these beams must display good shot-to-shot stability with high repetition rates, along with low emittance to ensure transverse coherence.\\
\noindent Recent advances have significantly improved the stability of LWFA beams. High-quality statistics have been demonstrated on a continuous $24\,$h experiment at $1\,$Hz in \cite{maier_decoding_2020}, and the continuous and stable operation of a kHz LWFA accelerator for $5\,$hours using a free-flowing $\mathrm{N_2}$ jet has been demonstrated in \cite{rovige_demonstration_2020}. Latest experiments confirmed the potential of LWFA for producing highly stable and well-defined low-energy electron beams at kHz, with beam-pointing fluctuations below 2.3 mrad RMS and charge fluctuations under 7\% RMS \cite{monzac_differential_2025}.\\

\noindent The quality of a particle beam can be quantified using its emittance, which describes the volume it occupies in the 6D phase-space defined by spatial coordinates ($x$, $y$, $z$) and momenta ($p_x$, $p_y$, $p_z$). The transverse emittances are projections of the beam’s phase-space distribution: for a beam propagating along $z$, the transverse normalized emittance along $x$ is a projection onto the ($x$, $p_x$) plane: $\epsilon_{n,x} = \frac{1}{m_ec}\sqrt{\langle x^2\rangle\langle p_x^2\rangle-\langle xp_x \rangle^2}$ with $m_e$ the electron mass and $c$ the speed of light. The transverse normalized emittance represents the beam's spatial quality and its ability to be tightly focused. In practice, experimental methods rather give access to the transverse geometric emittance $\epsilon_{g,x}=\sqrt{\langle x^2\rangle\langle x'^2\rangle-\langle xx' \rangle^2}$, a quantity defined in trace-space ($x$, $x’$) where $x’ = p_x/p_z$ is an angle. For a monochromatic beam, the geometric and normalized emittance are simply linked by: $\epsilon_{n,x} = \gamma\beta_z\epsilon_{g,x}$ where $\gamma$ is the Lorentz factor and $\gamma\beta_z = p_z/m_ec$ the normalized longitudinal momentum, so that the normalized emittance can be easily retrieved from the measurement of the geometric emittance.\\

\noindent The normalized transverse emittance $\epsilon_{n,x}$ of the electron beam directly affects its transverse coherence length $L_T$. For a beam with small divergence, $L_T$ at the sample position is given by \cite{van_oudheusden_electron_2007}:
\begin{equation}
    L_T = \frac{\hbar}{m_ec}\frac{\sigma_x}{\epsilon_{n,x}} 
\end{equation}
where $\sigma_x$ is the size of the beam on the sample. Moreover, to resolve interferences using a sample with lattice parameter $a$, the transverse coherence length must exceed $a$ ($L_T > a$). Given $\sigma_x = 200\,$\textmu m and a lattice parameter $a = 5\,$\AA \ (similar to silicon) this imposes $\epsilon_{n,x} < 154\,\mathrm{nm \cdot rad}$. Thus, achieving low-emittance beams is essential for high-quality UED.\\

\noindent Several methods exist to measure the emittance of an electron beam, many of which are detailed in \cite{mcdonald_methods_1989}. However, measuring the emittance of electron beams generated by LWFA presents additional challenges due to the beams’ shot-to-shot fluctuations and inherently broader energy spread (compared to conventional accelerators) \cite{cianchi_challenges_2013}. While emittance measurements have been successfully performed for LWFA beams in previous studies \cite{fritzler_emittance_2004, sears_emittance_2010, brunetti_low_2010, weingartner_ultralow_2012}, such measurements have not yet been demonstrated for the few-MeV, kHz-rate beams that are particularly relevant for UED.\\

\noindent In this work, we introduce a compact method based on the combination of a permanent magnet solenoid and a dipole magnet, to measure the emittance of few-MeV, kHz repetition-rate LWFA beams for UED, and we validate it through a diffraction experiment. This article is structured as follows: Section \ref{sec2} introduces the experimental methods and apparatus. The experimental results are presented in Section \ref{sec3}, highlighting the achievement of a low normalized emittance comparable to that of conventional accelerators in the same energy range, and thus meeting UED requirements. Then, Section \ref{sec4} reports on the successful use of the low emittance beams for obtaining clear diffraction images from a single-crystal silicon sample. Finally, Section \ref{sec5} concludes the paper.

\section{\label{sec2} Experimental methods}

\subsection{Laser wakefield accelerator}

\begin{figure}[h!]
    \centering
    \includegraphics[width=0.9\linewidth]{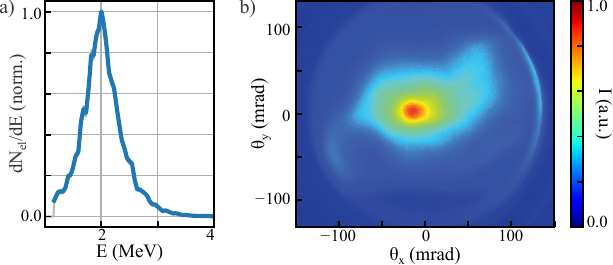}
    \caption{Typical spectrum (a) and beam profile (b) obtained during the experimental emittance measurement.}
    \label{fig:classic_diags}
\end{figure}

 The experimental setup is identical to that described in detail in \cite{monzac_optical_2024}, without implementing the differential pumping. The principal information is summarized below. The laser wakefield accelerator is driven by a kHz laser that delivers $4\,$fs pulses at FWHM with $2.5\,$mJ on-target energy \cite{ouille_relativistic-intensity_2020}. The pulses are focused by a $100\,$mm off-axis parabola down to a $4.5\,\text{\textmu}$m spot at FWHM, reaching a peak intensity in vacuum $I = 1.8 \times 10^{18}\,\mathrm{W \cdot cm}^{-2}$. The laser is focused at a distance of $150\,\text{\textmu}$m from the exit of a continuously flowing supersonic-shocked gas jet with a $180\,\text{\textmu}$m exit diameter \cite{tomkus_high-density_2018, marcinkevicius_femtosecond_2001} with $N_2$ gas. When performing the emittance measurement, the electron beam charge typically ranged from $200\,$fC to $1\,$pC, with a peak energy around $2\,$MeV. Figure \ref{fig:classic_diags} displays a typical spectrum and beam profile from this series of experimental measurements.

\begin{figure*}[t]
\includegraphics[width = 0.85\linewidth]{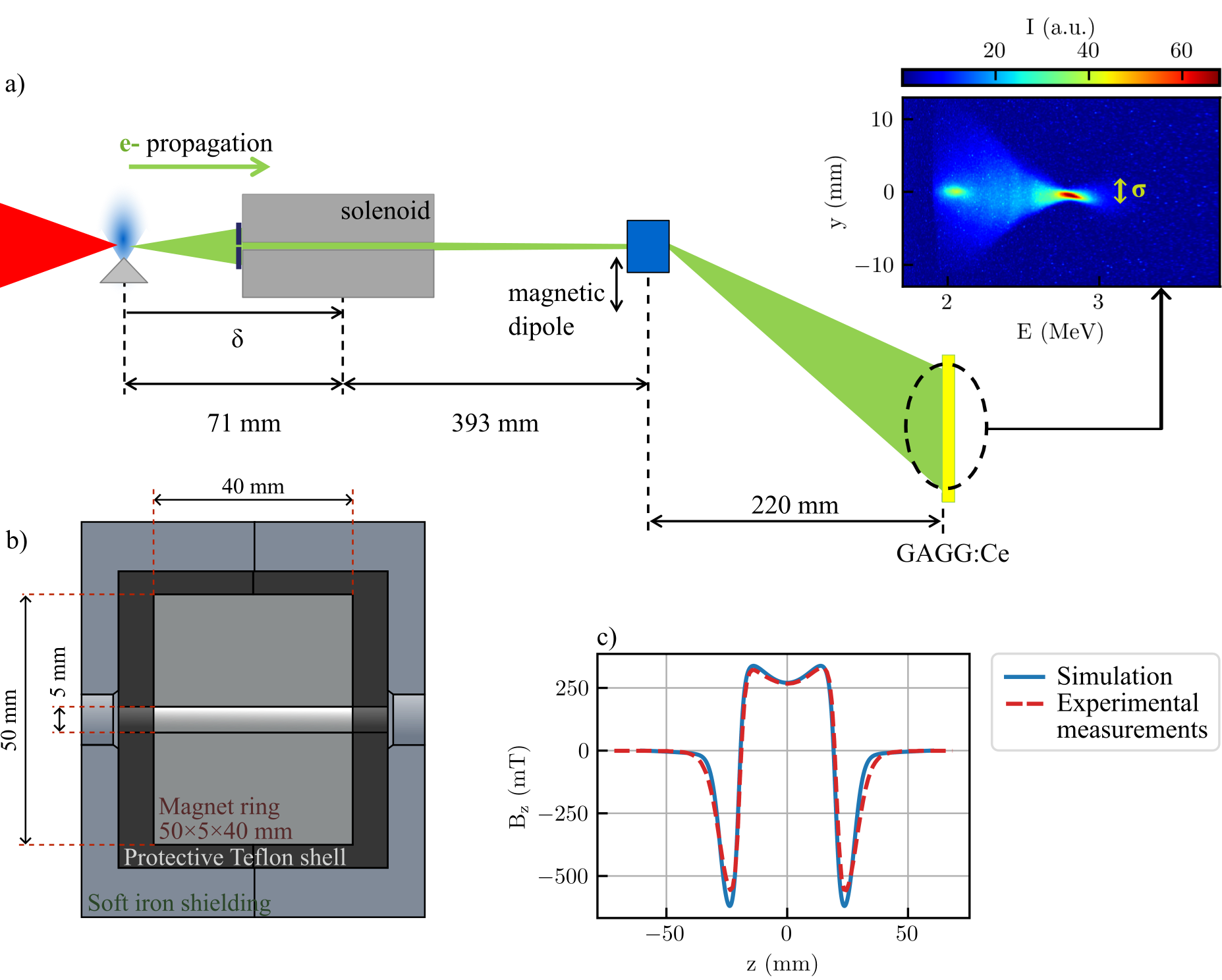}
\caption{\label{fig:setup} a) Experimental setup.  Electrons, accelerated by laser-wakefield acceleration, pass through the focusing solenoid and the dispersive dipole (which separates them by energy), then hit the scintillator. Without the dipole, the system has a magnification $M \simeq 8$.  b) Schematic of the permanent solenoid magnet enclosed in the Teflon shell and soft iron shielding.\\ 
c) Comparison of the measured on-axis magnetic field and the field simulated using COMSOL.}
\end{figure*}

\subsection{Emittance measurement apparatus}

One of the most widely used techniques to measure a beam transverse emittance is the quadrupole scan method (or solenoid scan method) \cite{mcdonald_methods_1989, carlsten_measuring_1993}. In this approach, a quadrupole magnet or a solenoid is placed in the beam path, with a screen positioned downstream to monitor the beam profile. By systematically varying the magnet’s strength and measuring the resulting electron beam size on the screen, one can fit the data to extract the beam’s emittance. To refocus the electron beam, we decided to use a solenoid. For a $2.5\,$MeV electron beam, we typically need a solenoid with an effective magnetic field $B_{eff} = 0.25\,$T and a length $L = 40\,$mm. Traditional current-based solenoids follow the formula:
\begin{equation}
B_{eff} = \frac{\mu_0NI}{L}
\end{equation}
\noindent where $\mu_0$ is the vacuum permeability, $I$ is the current through the coils, and $N$ is the number of turns in the coil. To achieve the desired magnetic field, we would need $NI = 8 \times 10^3\,$A. However, the solenoid must be placed as close as possible to the electron source, in vacuum, and the need for cooling makes traditional coils impractical. Instead, we use a custom permanent magnet solenoid which offers a more compact and cost-effective alternative.\\

\noindent The experimental setup for emittance measurement is illustrated in Figure \ref{fig:setup}a. Electrons are accelerated by a laser-plasma accelerator and then pass through a permanent magnet solenoid located at a distance $\delta$ from the gas jet, which is the electron source. Here $\delta$ is the distance between the gas jet and the solenoid center. An aluminum pinhole with diameter $d = 1\,$mm is placed before the solenoid in order to reduce the divergence of the beam and eliminate the influence of pointing jitter on our emittance measurement. Electrons with a specific energy of $2.5\,$MeV are focused by the solenoid onto a scintillator screen detector, placed $68.4\,$cm downstream of the gas jet. The RMS transverse size of the beam, $\sigma$, is observed on the scintillator for various positions of the solenoid $\delta$. The emittance of the beam is then retrieved by fitting the $\sigma(\delta)$ curve with General Particle Tracker simulations \cite{gpt} of electron beam trajectory (see Section \ref{sec:GPT_sim}).\\
\noindent\textit{Solenoid.} The magnet used to focus the electron beam is a custom NdFeB ring magnet with dimensions $20\times05\times40\,$mm. The magnetic field is confined using a soft iron shield (thickness $\simeq 10\,$mm), mounted around a protective Teflon shield (thickness $\simeq 5\,$mm). A schematic of the solenoid mounted in the shell is shown in Figure \ref{fig:setup}b where the different layers are identified. The soft iron has a magnetic permeability $\mu_r = 1000$, which allows it to effectively confine magnetic field lines. The solenoid’s magnetic field was simulated using COMSOL Multiphysics, with the simulations incorporating the detailed geometry of the magnets and shielding, with specified materials to replicate real-world conditions. The simulations were validated against experimental measurements, and Figure \ref{fig:setup}c compares the measured longitudinal on-axis magnetic field with the simulated results.\\
\noindent\textit{Detector.} The detector consists of a GAGG:Ce scintillator imaged on a ANDOR Neo 5.5 Megapixel camera. This camera introduces minimal uncertainty, equivalent to the signal from a single electron. Particularly low signal levels were anticipated, and this camera was used to enhance the signal-to-noise ratio. The GAGG:Ce scintillator has a high light yield ($60\,$photons/keV, approximately twice that of YAG:Ce), making it more efficient in converting radiation into visible light. To preserve spatial resolution, the scintillator is just $0.1\,$mm thick, and is mounted on a $1\,$mm UVFS substrate for structural support. The scintillator is protected by a reflective $\mathrm{BaSO_4}$ layer, coated with a thin $0.06\,$mm aluminum layer that acts as a filter for low-energy electrons. Using a USAF resolution target, the imaging system was characterized to resolve features as small as $15.6\,$\textmu m FWHM (corresponding to RMS size $\sigma = 6.6\,$\textmu m).\\

\begin{figure*}[t]
\includegraphics[width = 0.95\linewidth]{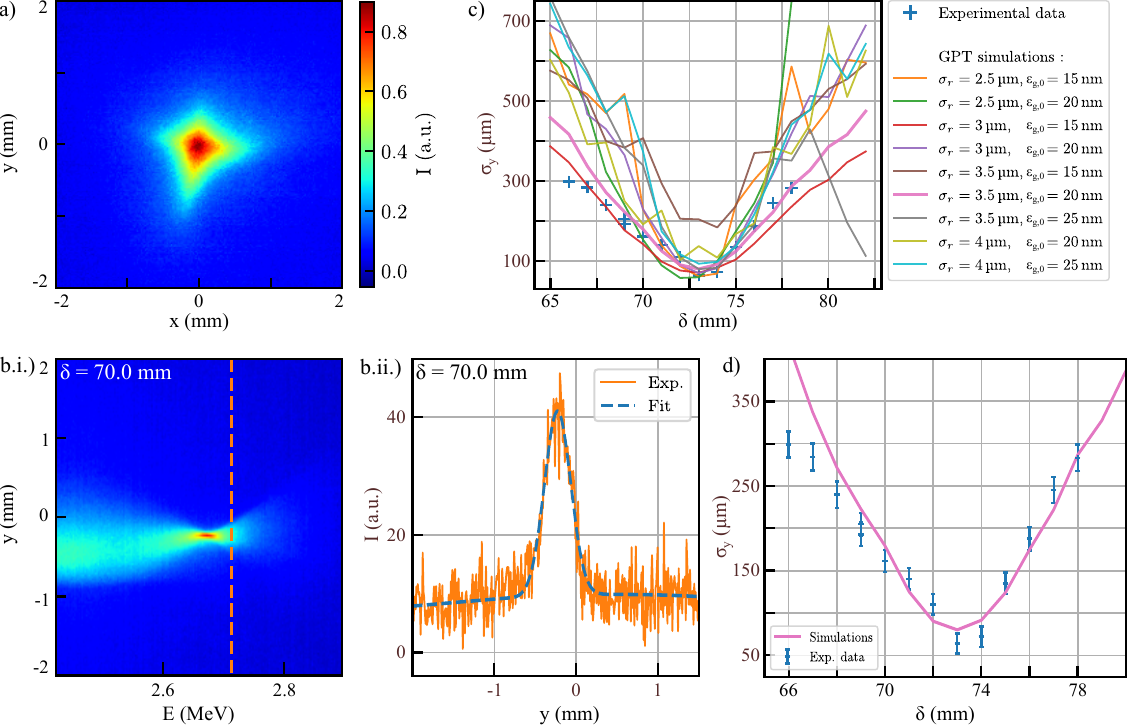}
\caption{\label{fig:exp_res} a) Image of the focal spot with no dispersive dipole. b) Experimental data of the emittance measurement for $\delta = 70.0\,$mm. b.i. shows the focused beam registered from the GAGG emission, the orange line indicates the energy $E = 2.7\,$MeV selected fo analysis and b.ii. shows the cross-section of the beam in the $y$ direction for this energy. c) Lines: simulated evolution of the measured cross-section beam size for E = $2.7\,$MeV as a function of the solenoid position for various sets of initial conditions. Blue crosses: experimental data for E = $2.7\,$MeV. d) Best fit of the experimental data with simulated data using initial parameters $\sigma_r = 3.5\,$\textmu m and $\epsilon_g = 20\,\mathrm{nm \cdot rad}$. The error bars account both for the RMS fluctuations of the experimental data and the resolution limits of the imaging system.}

\end{figure*}

\noindent \textbf{Overcoming the chromaticity problems}\\
\noindent The traditional solenoid scan method for emittance measurements is best suited for beams with a narrow energy spectrum \cite{carlsten_measuring_1993, sears_emittance_2010, setiniyaz_beam_2016, lindstrom_emittance_2024}. This is because a solenoid has an equivalent focal length $f$ given by:
\begin{equation}
    f = \frac{4m_e^2c^2(\gamma^2 - 1)}{e^2B_{eff}^2L}
\end{equation}
\noindent where $e$ is the elementary charge, $B_{eff}$ the solenoid's effective magnetic field, and $L$ its length. The Lorentz factor $\gamma$ is defined as $\gamma = 1 + E/m_ec^2$, where $E$ is the kinetic energy of the electrons. It is clear from this expression that the focal length varies quadratically with electron energy, making the solenoid a highly chromatic device. This chromatic behavior poses a significant experimental challenge: precise focusing requires a well-defined electron energy. However, LWFA-accelerated electron beams are rarely mono-energetic, particularly at low energies. For this experimental campaign, the electron beam has a peak energy of $2\,$MeV with a $0.8\,$MeV energy spread ($\Delta E / E = 40\%$). The beamline is designed to focus $2.5\,$MeV electrons onto the scintillator for $\delta = 65\,$mm. Consequently, electrons with energies below $2.5 \,$MeV are focused upstream of the scintillator, those with higher energies are focused downstream. This energy spread results in a blurred focal spot on the scintillator, complicating the measurement of the electron beam size.\\
\noindent Facing a similar issue, Weingartner \textit{et al.} proposed to disperse the electrons using a magnetic dipole placed in the electron beam path \cite{weingartner_ultralow_2012}. The role of the dipole is to separate the energies in the focal plane where the detector is positioned. The dispersed focus then displays a bow-tie shape, where the pinch point corresponds to the optimum focused energy, while other energies are out of focus. The experimental set-up displayed in Figure \ref{fig:setup}a includes the dispersive dipole and shows the bow-tie shape at the scintillator position, with the focus at $2.8\,$MeV. When varying the position $\delta$ of the solenoid, the width $\sigma$ of the beam on the detector for a specific energy can be monitored, allowing for more precise measurements.\\
\noindent\textit{Dispersive dipole.} The dispersive dipole is a set of two NdFeB magnets with dimensions $15 \times 15 \times 4\,$mm. These magnets are separated by a distance $d = 18\,$mm, generating a magnetic field of approximately $120\,$mT over $15\,$mm along the electron beam path. They are encased in a soft iron shielding, to prevent the magnetic field from affecting electron propagation when the dipole is not in the beam path. At $2.5\,$MeV, the energy resolution is $\delta E / E = 5\,\%$, ensuring sufficient precision for the emittance measurement.

\section{\label{sec3} Experimental results}

\subsection{Initial focusing configuration}

\noindent We first focus the electron beam onto the GAGG scintillator without the dispersive dipole, where electrons travel directly from the gas jet to the GAGG through the solenoid. Figure \ref{fig:exp_res}a presents an example of a measured focal spot. Using 20 acquisitions (each with an exposure time $e = 40\,$ms), the FWHM spot size is measured to be $\phi_x = 0.609 \pm 0.023\,\mathrm{mm}$ and $\phi_y = 0.615 \pm 0.034\,\mathrm{mm}$. The spot size exhibits minimal variation when the solenoid is moved longitudinally, due to the 40\% energy spread in the electron beam. This spread results in a blurred focal spot, as electrons of different energies are focused at various positions before and after the screen. Additionally, a cross-shaped pattern is observed, likely due to residual aberrations in the solenoid alignment. 

\subsection{Emittance measurement}

\noindent Next, the dispersive dipole is introduced in the beamline. In this configuration, the electrons first pass through the solenoid and are then dispersed by the dipole. The resulting beam is observed on the GAGG screen where the expected bow-tie-shaped distribution is observed. The solenoid's position is scanned from $\delta = 65\,$mm to $\delta = 78\,$mm, with the corresponding beam profiles being recorded at each position. Figure \ref{fig:exp_res}b shows the results for $\delta = 70\,$mm. The i. panel displays the GAGG emission image, representing the spatial electron beam profile. The specific energy E = $2.7\,$MeV, highlighted by the orange line, is selected for analysis. The beam cross-section at this energy is plotted in the ii. panel. The data are fitted to a function that is the sum of a fifth-degree polynomial and a Gaussian: 
\begin{equation}
    f(y) = a y^5 + b y^4 + c y^3 + d y^2 + e y + f + A \exp{\left(-\frac{(y - \mu)^2}{2 \sigma^2}\right)}
\end{equation}

\noindent The sigma value extracted from the fit is recorded for each solenoid position $\delta$. Figure \ref{fig:exp_res}c presents the $\sigma_y(\delta)$ curve for E = $2.7\,$MeV with the blue crosses. The beam cross-section at E = $2.7\,$MeV decreases to a minimum $\sigma_y = 64.2 \pm 12.1\,$\textmu m at $\delta = 73\,$mm before increasing again. The data point at $\delta = 69\,$mm is duplicated, as a control measurement was taken at this position at the end of the scan. For each solenoid position 20 images were acquired, each with an exposure time of $900\,$ms (corresponding to $900$ shots) for performing statistics. Note that due to the dispersion introduced by the dipole in the $x$ direction, only the beam size in the $y$ direction $\sigma_y$ is accessible, limiting the measurement to the transverse emittance in this direction. To measure the emittance in the $x$ direction, the dipole must be rotated.\\

\subsection{\label{sec:GPT_sim} General Particle Tracer simulations}

To retrieve the emittance from the measured $\sigma(\delta)$ curve while accounting for the real experimental magnetic field, numerical simulations of the electron beam propagation through the beamline were conducted using General Particle Tracer (GPT). GPT is a software package developed by Pulsar Physics to study the 3D dynamics of charged particles in electromagnetic fields \cite{gpt}.\\
\noindent \textit{Initialization of the simulations.} GPT simulations were performed using 50,000 macro-particles to model the electron beam, with a total charge $Q = 500\,$fC. Space-charge was not considered for these simulations after a benchmark test comparing results with and without space-charge showed no significant difference. The initial spatial distribution of electrons follows a centered Gaussian distribution, with an adjustable RMS width $\sigma_{r_{0}}$. The beam divergence also follows a Gaussian distribution with an RMS width $\sigma_{\theta_{xy}} = 25\,$mrad, but it is constrained further in the beamline by a pinhole at the solenoid entrance with diameter $d = 1.0\,$mm, as in the experiment. The beam geometric emittance $\epsilon_0$ is an adjustable parameter. The energy distribution follows a Gaussian distribution centered at $E = 2.7\,$MeV with an RMS energy spread $\sigma_E = 0.5\,$MeV RMS (negative energies were excluded as they are not physical).\\ 
\noindent \textit{Magnetic elements.} The beamline's magnetic field for each element, whether solenoid or dipole, was simulated using either Poisson Superfish (PSF) or COMSOL Multiphysics, and validated against experimental measurements. These simulations incorporated the detailed geometry of the magnets and shielding, with specified materials to closely replicate real-world conditions.\\
\noindent \textit{Reproducing the $\sigma(\delta)$ curve.} Electrons are generated at $z = 0$, and the screen (scintillator) is positioned at $z = 0.684\,$m to match experimental conditions. The positions of the electrons are recorded as they crossed the screen. Various initial beam sizes ($\sigma_{r_{0}}$) and initial emittances ($\epsilon_0$) were explored with the simulations. For each set of initial parameters ($\sigma_{r_{0}},\ \epsilon_0$), multiple simulations were performed by scanning the longitudinal position $\delta$ of the solenoid and the corresponding beam images on the screen were registered, mimicking the experimental procedure. The beam cross-section for the desired energy was extracted, and the $\sigma(\delta)$ curve was plotted, as in the experiments.\\
\noindent Residual aberrations in the experimental beamline (see Figure \ref{fig:exp_res}a) lead to an overestimation of the beam size at the focus in our measurements due to the imperfect focusing of the electron beam. These aberrations are not accounted for in the GPT simulations and thus the retrieved emittance is similarly overestimated: the measured emittance represents an upper bound for the actual emittance.

\begin{figure*}[t]
    \centering
    \includegraphics[width=0.78\linewidth]{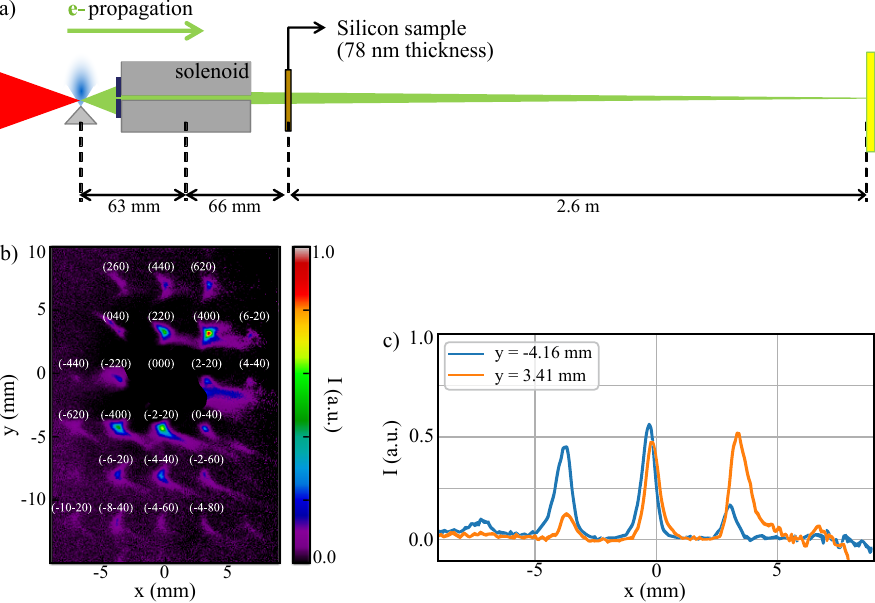}
    \caption{a) Experimental setup for electron diffraction. Electrons are accelerated via laser-wakefield acceleration and focused by the solenoid onto a scintillator. The silicon sample, on which the electrons diffract, is positioned immediately after the solenoid. The resulting diffraction pattern is observed on the scintillator. b) Diffraction pattern with the sample in place. The image is acquired with a $10\,$s exposure time, corresponding to $10^4$ shots. c) Cross-sectional analysis of the diffraction pattern for $y = 3.41\,$mm and $y = -4.16\,$mm.}
    \label{diff}
\end{figure*}

\subsection{Fitting experimental data with simulations}

Fine-tuned GPT simulations were performed for a range of initial beam sizes and emittances, and the simulated data were compared with the experimental results. The initial beam size $\sigma_{r}$ (standard deviation for a Gaussian distribution) was varied from $2.5$ to $4\,$\textmu m in steps of $0.5\,$\textmu m. The initial beam emittance $\epsilon_{g, 0}$ was tested from $15$ to $25\,\mathrm{nm\cdot rad}$ in $5\,\mathrm{nm\cdot rad}$ increments.\\
\noindent Figure \ref{fig:exp_res}c displays the simulated curves overlaid with the experimental data represented by the blue crosses. Among these, the simulation with $\sigma_{r} = 3.5\,$\textmu m and $\epsilon_{g,0} = 20\,\mathrm{nm \cdot rad}$ provides the best match to the experimental results, as shown in Figure \ref{fig:exp_res}d. The error bars for the experimental data account for the optical resolution and the uncertainties in fitting the data to determine the width of the peak. The excellent agreement between simulation and experiment leads to the conclusion that the geometric emittance of the beam is $\epsilon_g = 20\,\mathrm{nm \cdot rad}$. Given that this was found at the energy E = $2.7\,$MeV, the normalized emittance of the beam is $\epsilon_n = 124\,\mathrm{nm \cdot rad}$. This result was achieved using a spatial filtering of the beam, with 10\% of the initial electron beam charge being transmitted.

\section{\label{sec4} Application to electron diffraction}

The low transverse emittance measured ($\epsilon_n = 124\,\mathrm{nm \cdot rad}$) satisfies $\epsilon_{n,x} \leq 154\,\mathrm{nm \cdot rad}$ required for resolving diffraction patterns in materials with a lattice parameter  $a \simeq 5\,$\AA$\ $(see Section \ref{sec1}). To demonstrate this capability, an electron diffraction experiment was conducted using the available electron beam on a silicon sample.

\subsection{Setting up the diffraction experiment}

The layout of the experimental setup for the diffraction experiment is illustrated in Figure \ref{diff}a.\\
\noindent \textit{Sample.} A silicon sample consisting of a grid of a single-crystal silicon nanomembrane was used. The nanomembrane has a thickness of $78\,$nm. Silicon has a diamond cubic lattice structure with a lattice parameter $a = 5.431\,$\AA\ at $T = 300\,$K. The sample is mounted on a copper holder pierced with holes positioned $31\,$mm after the solenoid. The holder is attached to motorized translation stages that allow for alignment along the $x$ and $y$ axes, ensuring that the sample can be precisely aligned with the electron beam.\\
\noindent \textit{Detector.} To resolve the Bragg peaks on the detector, a separation of $d = 1\,$mm between the peaks at the detector position is desired. Using the Bragg formula, $n\lambda_B = 2a \sin \theta_B$ with $n \in \mathbb{Z}$, and assuming a $2.5\,$MeV electron beam ($\lambda_B = 0.42\,$pm), the separation angle between the peaks is $\theta_B = 0.38\,$mrad.  To achieve this peak separation of $1\,$mm, the screen must be positioned $2.6\,$m behind the sample. In this case the scintillator is a LANEX, and it is positioned $2.67\,$m behind the solenoid.\\

\vspace{1cm}

\subsection{Experimental results}

\noindent After inserting the sample in the beam path, the signal on the LANEX screen is optimized for intensity by fine-tuning the gas jet position (with position adjustments under $5\,$\textmu m) and the source-solenoid distance to maximize the signal-to-noise ratio. This optimization process may affect the main energy of the electron beam. Once optimized, a distinct diffraction pattern emerges on the LANEX screen. Figure \ref{diff}b. shows the resulting diffraction image. The 0-order peak has been attenuated to better visualize the Bragg peaks, which are clearly observable up to the 3rd order. The peaks are not perfectly circular, due to an imperfect alignment of the solenoid. A cross-sectional analysis of the diffraction pattern observed on the screen is shown in Figure \ref{diff}c. The intensity profiles along $y = 3.41\,$mm and $y = -4.16\,$mm are plotted. The distance between two adjacent peaks is approximately $3.5\,$mm.\\
\noindent Given that the screen is positioned $2.67\,$m behind the sample and the silicon sample has a known lattice parameter $a = 5.431\,$\AA, we can determine the actual wavelength of the electrons using the Bragg formula, which yields $\lambda_B = 1.46\,$pm. Using the de Broglie relation:
\begin{equation}
    \label{eq:Broglie_wl}
    \lambda_B = \frac{h}{m_ec\sqrt{\gamma^2 -1}}
\end{equation}

\noindent this wavelength corresponds to an electron energy of approximately $500\,$keV. As we optimized the image based on the signal intensity and since we could not simultaneously measure the electron spectrum and the diffraction pattern, it is probable that the electron beam energy decreased to 500 keV during the signal optimization process.

\subsection{Discussion}

\begin{figure}
    \centering
    \includegraphics[width=0.8\linewidth]{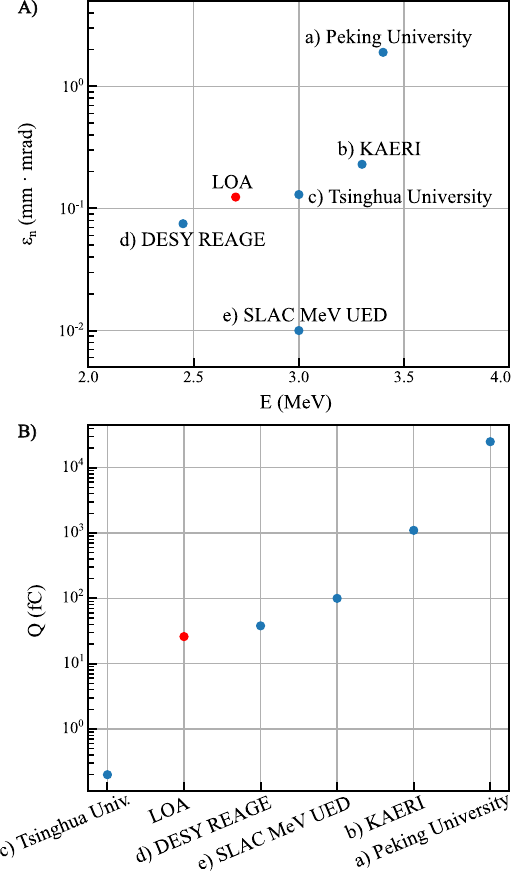}
    \caption{A) Review of the emittance reported by various conventional accelerator facilities with energies below $10\,$MeV. a) \cite{liu_experimental_2018}. b) \cite{setiniyaz_beam_2016}. c) \cite{li_experimental_2009}. d) \cite{desy_regae_2025}. e) \cite{mev-ued_2025}. B) Corresponding electron beam charges reported.}
    \label{review}
\end{figure}

The reported emittances for various radio-frequency electron guns around the world are displayed in Figure \ref{review} in blue. Our LWFA-based accelerator is highlighted in red. In the few MeV energy range, the best emittance is reported from the SLAC Ultrafast Electron Diffraction (UED) facility, with a value of $\epsilon_{n, SLAC} = 10\,\mathrm{nm \cdot rad}$. Other existing accelerators in this energy domain have emittances ranging from $\epsilon_n = 60\,\mathrm{nm \cdot rad}$ to $\epsilon_n = 1.9\,\mathrm{\text{\textmu}m \cdot rad}$. Those accelerators deliver electron bunches with charges spanning from $0.2\,$fC to $30\,$pC. In light of this, our accelerator's performance is notably competitive, with $\epsilon_n = 124\,\mathrm{nm \cdot rad}$ for a bunch charge $Q = 25 \pm 5\,$fC RMS. This demonstrates that kHz LWFA can achieve high beam quality (in terms of emittance), comparable to conventional accelerators.\\

\noindent The subsequent electron diffraction experiment confirmed the potential of our LWFA source for UED studies. With a $25\,$fC beam, we successfully observed diffraction from a silicon sample, with Bragg peaks clearly visible over several orders, despite a non-ideal focal spot and an electron energy of $E \simeq 500\,$keV. However, the non-monoenergetic nature of the beam degrades the temporal resolution due to bunch elongation during propagation. For a relativistic electron bunch, the duration increase after propagating a distance $L$ is given by \cite{he_capturing_2016, faure_concept_2016}:
\begin{equation}
    dt = \frac{L}{c}\frac{d\gamma}{\gamma}\frac{1}{\gamma^2}
\end{equation}
\noindent In the current setup, with the sample positioned $L = 129\,$mm downstream of the gas jet, an energy $E \simeq 500\,$keV and considering $d\gamma / \gamma = 20\%$, the bunch duration broadens by $dt = 22\,$ps between the source and the sample. If the diffraction experiment is performed using the best accelerator parameters reported in \cite{monzac_optical_2024} with $E = 7\,$MeV and $\Delta E / E = 10\%$, the broadening would be reduced to $\Delta t \approx 200\,$fs. Then to achieve sub-10 fs resolution two approaches can be considered. The first, proposed in \cite{he_capturing_2016}, involves streaking techniques, which convert temporal information into a spatial diagnostic using the linear energy chirp induced during the drift of the beam in free-space. The second idea is to develop an advanced electron beam transport line to recompress the bunch and minimize the effective energy spread, as demonstrated in \cite{faure_concept_2016}. In any case, more work is now needed to tackle the time resolution challenge with positive prospects for reaching sub-10 fs resolutions \cite{faure_concept_2016, he_capturing_2016}.\\

\noindent In parallel, it is important to highlight that state-of-the-art RF guns have achieved remarkable performance, producing electron bunches as short as $6\,$fs RMS with minimal jitter using terahertz streaking \cite{zhao_terahertz_2018}, or $25\,$fs RMS bunches with charges up to $0.6\,$pC and arrival time jitter as low as $7.8\,$fs \cite{kim_towards_2020}: these results set a high benchmark for comparison.

\section{\label{sec5} Conclusion}

The primary objective of this study was to produce and characterize an ultra-low emittance beam from a high-repetition rate LWFA and to evaluate the feasibility of conducting electron diffraction experiments in the MeV range using a laser-wakefield accelerated electron beam. The measured normalized transverse emittance, $\epsilon_{n,x} = 124\,\mathrm{nm \cdot rad}$, is comparable to that of conventional accelerators, demonstrating our accelerator’s potential to achieve high beam quality in terms of emittance.\\
\noindent Encouraged by these results, we conducted an electron diffraction experiment using the available electron beam. This experiment showcases the significant potential of our kHz LWFA source for electron diffraction studies. With a $25\,$fC beam, we successfully observed diffraction from a silicon sample, with Bragg peaks clearly visible over several orders despite the non-ideal focal spot. These results highlight the capability of the setup to resolve fine structural details.\\
\noindent With further optimization such as improving beam focusing and repeating the experiment with the highest-quality electron beam demonstrated with our accelerator, this platform could be used to test pump-probe experiments, paving the way for time-resolved structural studies using LWFA electron beams.\\

\begin{acknowledgments}
\noindent This project has received funding from the European Union's Horizon 2020 research and innovation program under grant agreement JRA PRISE no.\,871124 Laserlab-Europe and IFAST under Grant Agreement No 101004730. This project was also funded by the Agence Nationale de la Recherche under Contract No. ANR-20-CE92-0043-01, and benefited from the support of Institut Pierre Lamoure via the “Accélération laser-plasma haute cadence” fund.\\
\noindent We also acknowledge Laserlab-Europe, Grant No. H2020 EC-GA 654148 and the Lithuanian Research Council under Grant agreement No. S-MIP-21-3.\\

\end{acknowledgments}

\section*{Data Availability Statement}
The data that support the findings of this study are available from the corresponding author upon reasonable request.

\bibliography{bib_global}

@article{he_capturing_2016,
	title = {Capturing Structural Dynamics in Crystalline Silicon Using Chirped Electrons from a Laser Wakefield Accelerator},
	volume = {6},
	issn = {2045-2322},
	doi = {10.1038/srep36224},
	pages = {36224},
	number = {1},
	journal = {Scientific Reports},
	shortjournal = {Sci Rep},
	author = {He, Z.-H. and Beaurepaire, B. and Nees, J. A. and Gallé, G. and Scott, S. A. and Pérez, J. R. Sánchez and Lagally, M. G. and Krushelnick, K. and Thomas, A. G. R. and Faure, J.},
	date = {2016-11-08},
    year = {2016},
}

@article{rovige_demonstration_2020,
	title = {Demonstration of stable long-term operation of a kilohertz laser-plasma accelerator},
	volume = {23},
	doi = {10.1103/PhysRevAccelBeams.23.093401},
	pages = {093401},
	number = {9},
	journal = {Physical Review Accelerators and Beams},
	shortjournal = {Phys. Rev. Accel. Beams},
	author = {Rovige, L. and Huijts, J. and Andriyash, I. and Vernier, A. and Tomkus, V. and Girdauskas, V. and Raciukaitis, G. and Dudutis, J. and Stankevic, V. and Gecys, P. and Ouille, M. and Cheng, Z. and Lopez-Martens, R. and Faure, J.},
	date = {2020-09-14},
    year = {2020},
}

@article{maier_decoding_2020,
	title = {Decoding Sources of Energy Variability in a Laser-Plasma Accelerator},
	volume = {10},
	doi = {10.1103/PhysRevX.10.031039},
	pages = {031039},
	number = {3},
	journal = {Physical Review X},
	shortjournal = {Phys. Rev. X},
	author = {Maier, Andreas R. and Delbos, Niels M. and Eichner, Timo and Hübner, Lars and Jalas, Sören and Jeppe, Laurids and Jolly, Spencer W. and Kirchen, Manuel and Leroux, Vincent and Messner, Philipp and Schnepp, Matthias and Trunk, Maximilian and Walker, Paul A. and Werle, Christian and Winkler, Paul},
	date = {2020-08-18},
    year = {2020},
}

@article{oubrerie_controlled_2022,
	title = {Controlled acceleration of {GeV} electron beams in an all-optical plasma waveguide},
	volume = {11},
	doi = {10.1038/s41377-022-00862-0},
	pages = {180},
	journal = {Light: Science and Applications},
	author = {Oubrerie, Kosta and Leblanc, Adrien and Kononenko, Olena and Lahaye, Ronan and Andriyash, Igor A. and Gautier, Julien and Goddet, Jean-Philippe and Martelli, Lorenzo and Tafzi, Amar and Phuoc, Kim Ta and Smartsev, Slava and Thaury, Cedric},
	year = {2022},
}

@article{gonsalves_petawatt_2019,
	title = {Petawatt Laser Guiding and Electron Beam Acceleration to 8 {GeV} in a Laser-Heated Capillary Discharge Waveguide},
	volume = {122},
	doi = {10.1103/PhysRevLett.122.084801},
	pages = {084801},
	number = {8},
	journal = {Physical Review Letters},
	shortjournal = {Phys. Rev. Lett.},
	author = {Gonsalves, A.J. and Nakamura, K. and Daniels, J. and Benedetti, C. and Pieronek, C. and de Raadt, T.C.H. and Steinke, S. and Bin, J.H. and Bulanov, S.S. and van Tilborg, J. and Geddes, C.G.R. and Schroeder, C.B. and Toth, Cs. and Esarey, E. and Swanson, K. and Fan-Chiang, L. and Bagdasarov, G. and Bobrova, N. and Gasilov, V. and Korn, G. and Sasorov, P. and Leemans, W.P.},
	date = {2019-02-25},
    year = {2019}
}

@article{aniculaesei_acceleration_2023,
	title = {The acceleration of a high-charge electron bunch to 10 {GeV} in a 10-cm nanoparticle-assisted wakefield accelerator},
	volume = {9},
	issn = {2468-2047},
	doi = {10.1063/5.0161687},
	pages = {014001},
	number = {1},
	journal = {Matter and Radiation at Extremes},
	shortjournal = {Matter and Radiation at Extremes},
	author = {Aniculaesei, Constantin and Ha, Thanh and Yoffe, Samuel and Labun, Lance and Milton, Stephen and {McCary}, Edward and Spinks, Michael M. and Quevedo, Hernan J. and Labun, Ou Z. and Sain, Ritwik and Hannasch, Andrea and Zgadzaj, Rafal and Pagano, Isabella and Franco-Altamirano, Jose A. and Ringuette, Martin L. and Gaul, Erhart and Luedtke, Scott V. and Tiwari, Ganesh and Ersfeld, Bernhard and Brunetti, Enrico and Ruhl, Hartmut and Ditmire, Todd and Bruce, Sandra and Donovan, Michael E. and Downer, Michael C. and Jaroszynski, Dino A. and Hegelich, Bjorn Manuel},
	date = {2023-11-15},
    year = {2023},
}

@article{tajima_laser_1979,
	title = {Laser Electron Accelerator},
	volume = {43},
	issn = {0031-9007},
	doi = {10.1103/PhysRevLett.43.267},
	pages = {267--270},
	number = {4},
	journal = {Physical Review Letters},
	shortjournal = {Phys. Rev. Lett.},
	author = {Tajima, T. and Dawson, J. M.},
	date = {1979-07-23},
    year = {1979},
}

@article{sciaini_femtosecond_2011,
	title = {Femtosecond electron diffraction: heralding the era of atomically resolved dynamics},
	volume = {74},
	issn = {0034-4885},
	doi = {10.1088/0034-4885/74/9/096101},
	shorttitle = {Femtosecond electron diffraction},
	pages = {096101},
	number = {9},
	journal = {Reports on Progress in Physics},
	shortjournal = {Rep. Prog. Phys.},
	author = {Sciaini, Germán and Miller, R J Dwayne},
	date = {2011-08},
    year = {2011},
}

@article{kim_towards_2020,
	title = {Towards jitter-free ultrafast electron diffraction technology},
	volume = {14},
	issn = {1749-4893},
	doi = {10.1038/s41566-019-0566-4},
	pages = {245--249},
	number = {4},
	journal = {Nature Photonics},
	shortjournal = {Nat. Photonics},
	author = {Kim, Hyun Woo and Vinokurov, Nikolay A. and Baek, In Hyung and Oang, Key Young and Kim, Mi Hye and Kim, Young Chan and Jang, Kyu-Ha and Lee, Kitae and Park, Seong Hee and Park, Sunjeong and Shin, Junho and Kim, Jungwon and Rotermund, Fabian and Cho, Sunglae and Feurer, Thomas and Jeong, Young Uk},
	date = {2020-04},
    year = {2020},
	}

@article{salehi_laser-accelerated_2021,
	title = {Laser-Accelerated, Low-Divergence 15-{MeV} Quasimonoenergetic Electron Bunches at 1 {kHz}},
	volume = {11},
	doi = {10.1103/PhysRevX.11.021055},
	pages = {021055},
	number = {2},
	journal = {Physical Review X},
	shortjournal = {Phys. Rev. X},
	author = {Salehi, F. and Le, M. and Railing, L. and Kolesik, M. and Milchberg, H.M.},
	date = {2021-06-11},
    year = {2021},
}

@article{kim_stable_2017,
	title = {Stable multi-{GeV} electron accelerator driven by waveform-controlled {PW} laser pulses},
	volume = {7},
	issn = {2045-2322},
	doi = {10.1038/s41598-017-09267-1},
	pages = {10203},
	number = {1},
	journal = {Scientific Reports},
	shortjournal = {Sci Rep},
	author = {Kim, Hyung Taek and Pathak, V. B. and Hong Pae, Ki and Lifschitz, A. and Sylla, F. and Shin, Jung Hun and Hojbota, C. and Lee, Seong Ku and Sung, Jae Hee and Lee, Hwang Woon and Guillaume, E. and Thaury, C. and Nakajima, Kazuhisa and Vieira, J. and Silva, L. O. and Malka, V. and Nam, Chang Hee},
	date = {2017-08-31},
    year = {2017},
}

@article{miao_multi-gev_2022,
	title = {Multi-{GeV} Electron Bunches from an All-Optical Laser Wakefield Accelerator},
	volume = {12},
	doi = {10.1103/PhysRevX.12.031038},
	pages = {031038},
	number = {3},
	journal = {Physical Review X},
	shortjournal = {Phys. Rev. X},
	author = {Miao, B. and Shrock, J.E. and Feder, L. and Hollinger, R.C. and Morrison, J. and Nedbailo, R. and Picksley, A. and Song, H. and Wang, S. and Rocca, J.J. and Milchberg, H.M.},
	date = {2022-09-16},
    year = {2022},
}

@article{ouille_relativistic-intensity_2020,
	title = {Relativistic-intensity near-single-cycle light waveforms at {kHz} repetition rate},
	volume = {9},
	issn = {2047-7538},
	doi = {10.1038/s41377-020-0280-5},
	pages = {47},
	number = {1},
	journal = {Light: Science and Applications},
	shortjournal = {Light Sci Appl},
	author = {Ouillé, Marie and Vernier, Aline and Böhle, Frederik and Bocoum, Maïmouna and Jullien, Aurélie and Lozano, Magali and Rousseau, Jean-Philippe and Cheng, Zhao and Gustas, Dominykas and Blumenstein, Andreas and Simon, Peter and Haessler, Stefan and Faure, Jérôme and Nagy, Tamas and Lopez-Martens, Rodrigo},
	urldate = {2024-03-27},
	date = {2020-03-23},
    year = {2020},
}

@article{faure_concept_2016,
	title = {Concept of a laser-plasma-based electron source for sub-10-fs electron diffraction},
	volume = {19},
	doi = {10.1103/PhysRevAccelBeams.19.021302},
	pages = {021302},
	number = {2},
	journal = {Physical Review Accelerators and Beams},
	shortjournal = {Phys. Rev. Accel. Beams},
	author = {Faure, J. and van der Geer, B. and Beaurepaire, B. and Gallé, G. and Vernier, A. and Lifschitz, A.},
	date = {2016-02-19},
    year = {2016},
}

@article{marcinkevicius_femtosecond_2001,
	title = {Femtosecond laser-assisted three-dimensional microfabrication in silica},
	volume = {26},
	issn = {1539-4794},
	doi = {10.1364/OL.26.000277},
	pages = {277--279},
	number = {5},
	journal = {Optics Letters},
	shortjournal = {Opt. Lett., {OL}},
	author = {Marcinkevičius, Andrius and Juodkazis, Saulius and Watanabe, Mitsuru and Miwa, Masafumi and Matsuo, Shigeki and Misawa, Hiroaki and Nishii, Junji},
	date = {2001-03-01},
    year = {2001},
}

@article{tomkus_high-density_2018,
	title = {High-density gas capillary nozzles manufactured by hybrid 3D laser machining technique from fused silica},
	volume = {26},
	issn = {1094-4087},
	doi = {10.1364/OE.26.027965},
	pages = {27965--27977},
	number = {21},
	journal = {Optics Express},
	shortjournal = {Opt. Express, {OE}},
	author = {Tomkus, Vidmantas and Girdauskas, Valdas and Dudutis, Juozas and Gečys, Paulius and Stankevič, Valdemar and Račiukaitis, Gediminas},
	date = {2018-10-15},
    year = {2018},
}

@article{monzac_optical_2024,
	title = {Optical ionization effects in {kHz} laser wakefield acceleration with few-cycle pulses},
	volume = {6},
	doi = {10.1103/PhysRevResearch.6.043099},
	pages = {043099},
	number = {4},
	journal = {Phys. Rev. Res.},
	author = {Monzac, J. and Smartsev, S. and Huijts, J. and Rovige, L. and Andriyash, I. A. and Vernier, A. and Tomkus, V. and Girdauskas, V. and Raciukaitis, G. and Mackevičiūtė, M. and Stankevic, V. and Cavagna, A. and Kaur, J. and Kalouguine, A. and Lopez-Martens, R. and Faure, J.},
	year = {2024},
}

@article{monzac_differential_2025,
	title = {Differential pumping for {kHz} operation of a laser wakefield accelerator based on a continuously flowing hydrogen gas jet},
	volume = {96},
	issn = {0034-6748},
	doi = {10.1063/5.0246912},
	pages = {043302},
	number = {4},
	journal = {Review of Scientific Instruments},
	shortjournal = {Review of Scientific Instruments},
	author = {Monzac, J. and Smartsev, S. and Huijts, J. and Rovige, L. and Andriyash, I. A. and Vernier, A. and Tomkus, V. and Girdauskas, V. and Raciukaitis, G. and Mackevičiūtė, M. and Stankevic, V. and Cavagna, A. and Kaur, J. and Kalouguine, A. and Lopez-Martens, R. and Faure, J.},
	date = {2025-04-01},
    year = {2025},
}

@article{sears_emittance_2010,
	title = {Emittance and divergence of laser wakefield accelerated electrons},
	volume = {13},
	doi = {10.1103/PhysRevSTAB.13.092803},
	number = {9},
	journal = {Physical Review Special Topics - Accelerators and Beams},
	shortjournal = {Phys. Rev. {ST} Accel. Beams},
	author = {Sears, Christopher M. S. and Buck, Alexander and Schmid, Karl and Mikhailova, Julia and Krausz, Ferenc and Veisz, Laszlo},
	date = {2010-09-22},
    year = {2010},
}

@article{weingartner_ultralow_2012,
	title = {Ultralow emittance electron beams from a laser-wakefield accelerator},
	volume = {15},
	doi = {10.1103/PhysRevSTAB.15.111302},
	pages = {111302},
	number = {11},
	journal = {Physical Review Special Topics - Accelerators and Beams},
	shortjournal = {Phys. Rev. {ST} Accel. Beams},
	author = {Weingartner, R. and Raith, S. and Popp, A. and Chou, S. and Wenz, J. and Khrennikov, K. and Heigoldt, M. and Maier, A. R. and Kajumba, N. and Fuchs, M. and Zeitler, B. and Krausz, F. and Karsch, S. and Grüner, F.},
	date = {2012-11-05},
    year = {2012},
}

@inproceedings{mcdonald_methods_1989,
	location = {Berlin, Heidelberg},
	title = {Methods of emittance measurement},
	isbn = {978-3-540-46716-8},
	doi = {10.1007/BFb0018284},
	pages = {122--132},
	booktitle = {Frontiers of Particle Beams; Observation, Diagnosis and Correction},
	publisher = {Springer},
	author = {{McDonald}, K. T. and Russell, D. P.},
	editor = {Month, M. and Turner, S.},
	year = {1989},
}

@article{lindstrom_emittance_2024,
	title = {Emittance preservation in a plasma-wakefield accelerator},
	volume = {15},
	rights = {2024 The Author(s)},
	issn = {2041-1723},
	doi = {10.1038/s41467-024-50320-1},
	number = {1},
	journal = {Nature Communications},
	shortjournal = {Nat Commun},
	author = {Lindstrøm, C. A. and Beinortaitė, J. and Björklund Svensson, J. and Boulton, L. and Chappell, J. and Diederichs, S. and Foster, B. and Garland, J. M. and González Caminal, P. and Loisch, G. and Peña, F. and Schröder, S. and Thévenet, M. and Wesch, S. and Wing, M. and Wood, J. C. and D’Arcy, R. and Osterhoff, J.},
	date = {2024-07-19},
    year = {2024},
}

@article{setiniyaz_beam_2016,
	title = {Beam characterization at the {KAERI} {UED} beamline},
	volume = {69},
	issn = {0374-4884, 1976-8524},
	doi = {10.3938/jkps.69.1019},
	pages = {1019--1024},
	number = {6},
	journal = {Journal of the Korean Physical Society},
	shortjournal = {Journal of the Korean Physical Society},
	author = {Setiniyaz, Sadiq and Kim, Hyun Woo and Baek, In-Hyung and Nam, Jinhee and Chae, {MoonSik} and Han, Byung-Heon and Gudkov, Boris and Jang, Kyu Ha and Park, Sunjeong and Jeong, Young Uk and Miginsky, Sergey and Vinokurov, Nikolay},
	date = {2016-09},
    year = {2016},
}

@article{carlsten_measuring_1993,
	title = {Measuring emittance of nonthermalized electron beams from photoinjectors},
	volume = {331},
	rights = {https://www.elsevier.com/tdm/userlicense/1.0/},
	issn = {01689002},
	doi = {10.1016/0168-9002(93)90159-F},
	pages = {791--796},
	number = {1},
	journal = {Nuclear Instruments and Methods in Physics Research Section A: Accelerators, Spectrometers, Detectors and Associated Equipment},
	shortjournal = {Nuclear Instruments and Methods in Physics Research Section A: Accelerators, Spectrometers, Detectors and Associated Equipment},
	author = {Carlsten, Bruce E. and Goldstein, John C. and O'Shea, Patrick G. and Pitcher, Eric J.},
	date = {1993-07},
    year = {1993},
}

@article{van_oudheusden_electron_2007,
	title = {Electron source concept for single-shot sub-100 fs electron diffraction in the 100 {keV} range},
	volume = {102},
	issn = {0021-8979},
	doi = {10.1063/1.2801027},
	pages = {093501},
	number = {9},
	journal = {Journal of Applied Physics},
	shortjournal = {Journal of Applied Physics},
	author = {van Oudheusden, T. and de Jong, E. F. and van der Geer, S. B. and ’t Root, W. P. E. M. Op and Luiten, O. J. and Siwick, B. J.},
	date = {2007-11-01},
    year = {2007},
}

@article{cianchi_challenges_2013,
	title = {Challenges in plasma and laser wakefield accelerated beams diagnostic},
	volume = {720},
	issn = {0168-9002},
	doi = {10.1016/j.nima.2012.12.012},
	series = {Selected papers from the 2nd International Conference Frontiers in Diagnostic Technologies ({ICFDT}2)},
	pages = {153--156},
	journal = {Nuclear Instruments and Methods in Physics Research Section A: Accelerators, Spectrometers, Detectors and Associated Equipment},
	shortjournal = {Nuclear Instruments and Methods in Physics Research Section A: Accelerators, Spectrometers, Detectors and Associated Equipment},
	author = {Cianchi, A. and Anania, M. P. and Bellaveglia, M. and Castellano, M. and Chiadroni, E. and Ferrario, M. and Gatti, G. and Marchetti, B. and Mostacci, A. and Pompili, R. and Ronsivalle, C. and Rossi, A. R. and Serafini, L.},
	date = {2013-08-21},
    year = {2013},
}

@article{brunetti_low_2010,
	title = {Low Emittance, High Brilliance Relativistic Electron Beams from a Laser-Plasma Accelerator},
	volume = {105},
	doi = {10.1103/PhysRevLett.105.215007},
	pages = {215007},
	number = {21},
	journal = {Physical Review Letters},
	shortjournal = {Phys. Rev. Lett.},
	author = {Brunetti, E. and Shanks, R. P. and Manahan, G. G. and Islam, M. R. and Ersfeld, B. and Anania, M. P. and Cipiccia, S. and Issac, R. C. and Raj, G. and Vieux, G. and Welsh, G. H. and Wiggins, S. M. and Jaroszynski, D. A.},
	date = {2010-11-19},
    year = {2010},
}

@article{fritzler_emittance_2004,
	title = {Emittance Measurements of a Laser-Wakefield-Accelerated Electron Beam},
	volume = {92},
	doi = {10.1103/PhysRevLett.92.165006},
	pages = {165006},
	number = {16},
	journal = {Physical Review Letters},
	shortjournal = {Phys. Rev. Lett.},
	author = {Fritzler, S. and Lefebvre, E. and Malka, V. and Burgy, F. and Dangor, A. E. and Krushelnick, K. and Mangles, S. P. D. and Najmudin, Z. and Rousseau, J.-P. and Walton, B.},
	date = {2004-04-23},
    year = {2004},
}

@misc{gpt,
    title = {General Particle Tracer},
    howpublished  = {Pulsar Physics},
    url = {www.pulsar.nl/gpt},
    year = {accessed in 2024}
}

@article{liu_experimental_2018,
	title = {Experimental Study on the Electron Superconducting Linac and its Application},
	volume = {{SAP}2017},
	doi = {10.18429/JACOW-SAP2017-MOBH1},
	pages = {6 pages, 1.460 MB},
	journal = {Proceedings of the 13th Symposium on Accelerator Physics},
	author = {Liu, Kexin and Chen, Jia-Er and Feng, Liwen and Hao, Jiankui and Huang, Senlin and Lin, Lin and Quan, Shengwen and Wang, Fang and Xie, Huamu and Yang, Limin and Zhu, Feng},
	year = {2018},
}

@article{li_experimental_2009,
	title = {Experimental demonstration of high quality {MeV} ultrafast electron diffraction},
	volume = {80},
	issn = {0034-6748},
	doi = {10.1063/1.3194047},
	pages = {083303},
	number = {8},
	journal = {Review of Scientific Instruments},
	shortjournal = {Review of Scientific Instruments},
	author = {Li, Renkai and Tang, Chuanxiang and Du, Yingchao and Huang, Wenhui and Du, Qiang and Shi, Jiaru and Yan, Lixin and Wang, Xijie},
	date = {2009-08-18},
    year = {2009},
}

@misc{mev-ued_2025,
	title = {{MeV}-{UED} Specifications {\textbar} Linac Coherent Light Source},
	url = {https://lcls.slac.stanford.edu/instruments/mev-ued/specifications},
	urldate = {2025-07-03},
    year = {Accessed in March 2025},
}

@misc{desy_regae_2025,
	title = {{DESY} {REGAE} parameters},
	url = {https://regae.desy.de/regae\_accelerator/},
    urldate = {2025-07-03},
    year = {Accessed March 2025},
}

@article{zhao_terahertz_2018,
	title = {Terahertz Streaking of Few-Femtosecond Relativistic Electron Beams},
	volume = {8},
	doi = {10.1103/PhysRevX.8.021061},
	pages = {021061},
	number = {2},
	journal = {Physical Review X},
	shortjournal = {Phys. Rev. X},
	author = {Zhao, Lingrong and Wang, Zhe and Lu, Chao and Wang, Rui and Hu, Cheng and Wang, Peng and Qi, Jia and Jiang, Tao and Liu, Shengguang and Ma, Zhuoran and Qi, Fengfeng and Zhu, Pengfei and Cheng, Ya and Shi, Zhiwen and Shi, Yanchao and Song, Wei and Zhu, Xiaoxin and Shi, Jiaru and Wang, Yingxin and Yan, Lixin and Zhu, Liguo and Xiang, Dao and Zhang, Jie},
	date = {2018-06-08},
    year = {2018},
}

@article{lundh_few_2011,
	title = {Few femtosecond, few kiloampere electron bunch produced by a laser–plasma accelerator},
	volume = {7},
	issn = {1745-2473, 1745-2481},
	doi = {10.1038/nphys1872},
	pages = {219--222},
	number = {3},
	journal = {Nature Physics},
	shortjournal = {Nature Phys},
	author = {Lundh, O. and Lim, J. and Rechatin, C. and Ammoura, L. and Ben-Ismaïl, A. and Davoine, X. and Gallot, G. and Goddet, J-P. and Lefebvre, E. and Malka, V. and Faure, J.},
	date = {2011-03},
    year = {2011},
}

@article{van_Oudheusden_compression_2010,
  title = {Compression of Subrelativistic Space-Charge-Dominated Electron Bunches for Single-Shot Femtosecond Electron Diffraction},
  author = {van Oudheusden, T. and Pasmans, P. L. E. M. and van der Geer, S. B. and de Loos, M. J. and van der Wiel, M. J. and Luiten, O. J.},
  journal = {Phys. Rev. Lett.},
  volume = {105},
  issue = {26},
  pages = {264801},
  numpages = {4},
  year = {2010},
  month = {Dec},
  publisher = {American Physical Society},
  doi = {10.1103/PhysRevLett.105.264801},
  url = {https://link.aps.org/doi/10.1103/PhysRevLett.105.264801}
}

@article{musumeci_laser-induced_2010,
  title = {Laser-induced melting of a single crystal gold sample by time-resolved ultrafast relativistic electron diffraction},
  author = {Musumeci, P. and Moody, J. T. and Scoby, C. M. and Gutierrez, M. S. and Westfall, M.},
  journal = {Appl. Phys. Lett.},
  volume = {97},
  issue = {6},
  pages = {063502},
  year = {2010},
  month = {Dec},
  url = {https://doi.org/10.1063/1.3478005}
}

@article{miller_femtosecond_2014,
author = {R. J. Dwayne Miller },
title = {Femtosecond Crystallography with Ultrabright Electrons and X-rays: Capturing Chemistry in Action},
journal = {Science},
volume = {343},
number = {6175},
pages = {1108-1116},
year = {2014},
doi = {10.1126/science.1248488},
URL = {https://www.science.org/doi/abs/10.1126/science.1248488},
}

@book{zewail_4D_2009,
    author = {Zewail, A. H. and Thomas, J. M. },
    title = {4D Electron Microscopy},
    publisher = {Imperial College Press, London},
    year = {2009},
}

\end{document}